\documentclass[12pt]{article}
\usepackage{amsmath}
\usepackage{amssymb}
\usepackage{amsthm}
\usepackage{amsfonts}
\usepackage{graphicx}
\usepackage{textcomp}
\usepackage{color}
\usepackage{hyperref}
\usepackage{tikz-cd}
\numberwithin{equation}{section}

\begin{document}

\begin{titlepage}

\title{Weyl Covariant Quadratic Curvature Gravity in 3-Dimensional Riemann-Cartan-Weyl Space-Times}

\author{ Tekin Dereli\footnote{tdereli@ku.edu.tr}, Cem Yeti\c{s}mi\c{s}o\u{g}lu\footnote{cyetismisoglu@ku.edu.tr} \\ {\small Department of Physics, Ko\c{c} University, 34450 Sar{\i}yer, \.{I}stanbul, Turkey }}

\date{27 June 2019}

\maketitle

\begin{abstract}

\noindent We discuss locally Weyl (scale) covariant generalisation of quadratic curvature gravity theory in three dimensions using Riemann-Cartan-Weyl space-times. We show that this procedure of Weyl gauging yields a consistent generalisation for a particular class of quadratic curvature gravity theories which includes the New Massive Gravity theory. 
\end{abstract}

\vskip 2cm


\thispagestyle{empty}

\end{titlepage}

\maketitle			
\clearpage 

\section{Introduction}
\setcounter{page}{1}

\noindent Locally scale covariant theories are important because they allow the construction of high and low energy complete theories. In the context of gravitational theories, this aspect is very important due to lack of having a well behaved perturbative quantum gravity. To gain insight in the quantum gravity problem, three dimensional (3D) toy models \cite{leutwyler} provide us with important examples: Topologically massive gravity (TMG) \cite{deser-jackiw-templeton1, deser-jackiw-templeton2}, minimal massive gravity (MMG) \cite{bergshoeff et al3, baykal} and new massive gravity (NMG) \cite{bergshoeff-hohm-townsend1, bergshoeff-hohm-townsend2}, all of which are extensions of cosmological general relativity (GR) in three dimensions (3D). Because GR in 3D has no propagating degree of freedom \cite{deser-jackiw-tHooft}, these models are obtained by augmenting the GR action with extra terms such as coupling a vector Chern-Simons term (for TMG and MMG) or a particular combination of quadratic curvature invariants (for NMG). An important common feature of all of these models is that, when linearised around a 3D anti-de Sitter background, they yield a propagating massive spin-2 field, i.e. a graviton. \\ 

\noindent In our recent works, we studied the locally Weyl covariant TMG \cite{dereli-yetismisoglu3} and MMG theories \cite{dereli-yetismisoglu4} where, we used the powerful language of differential forms on Riemann-Cartan-Weyl (RCW) space-times. RCW space-times provide a natural, geometrical framework to discuss locally scale covariant theories \cite{dereli-tucker}. Choosing a specific non-metricity tensor and identifying it with the Weyl connection helps us provide a geometrical origin to scale transformations. In this geometrical framework, the scale covariant theories are defined in terms of a locally scale invariant action with a first order variational formalism. \\

\noindent In this paper, we study the locally Weyl covariant generalisation of quadratic curvature gravity (QCG) in 3D and the consistency of this procedure. QCG is defined via the action of 3D GR augmented with generic quadratic curvature invariants. This model is motivated by NMG and the new improved massive gravity (NIMG) \cite{dereli-yetismisoglu1, dereli-yetismisoglu2} models. On the one hand, in NIMG we have studied a general 3D gravitational model that contains TMG, MMG and NMG as subcases and obtained exact background solutions. NIMG action also contains the most general quadratic curvature invariants in its action. On the other hand, NMG is a parity invariant extension of GR given by a specific combination of the squares of Ricci tensor and scalar curvature. It is shown to admit TMG theory as its square root and at the linearised level it is equivalent to the unitary Pauli-Fierz theory for a massive spin-2 field. These properties make it desirable to discuss Weyl covariant extensions of NMG and NIMG. To cover both cases, we study the Weyl covariant generalisation of QCG. We show that, for a certain subset of our coupling parameters (which also cover the NMG theory), this generalisation is consistent. \\

\noindent The organisation of the paper is as follows. In section two, we discuss RCW space-times within the context of local scale transformations, and explain how we implement Weyl covariance in a gravitational theory. Then we move on to present the Lagrangian formulation of quadratic curvature gravity and its Weyl covariant extension. We check the consistency of this procedure in section three. Concluding remarks make up the fourth section. The technical details regarding the quadratic curvature invariants and derivation of equations are in the appendices A and B, respectively.

\section{Riemann-Cartan-Weyl Space-Times}

\noindent We follow the same conventions as in references \cite{dereli-yetismisoglu3, dereli-yetismisoglu4}, however, to set the notation we briefly explain RCW space-times in general. A RCW space-time is a triplet $\{M, g, \nabla\}$ where $M$ is a smooth $n$-manifold, $g$ is a non-degenerate, Lorentzian metric tensor on $M$, and  $\nabla$ is a linear connection on $M$. With the help of $g$-orthonormal frames $\{X_a\}$ and their dual co-frames $\{e^a\}$ defined via $e^a(X_b)=\iota_b e^a = \delta^a_b$, the metric tensor can be expressed as $g=\eta_{ab} e^a \otimes e^b$ where $\eta_{ab} = g(X_a,X_b)= (-,+,..,+)$. For brevity, we use short hand notations for the exterior products $e^{ab\dots}\equiv e^a \wedge e^b \wedge \dots $, and the interior products $\iota_{ab\dots} \equiv \iota_a \iota_b \dots$. The metric allows the definition of the Hodge duality operator $*:\Lambda^p(M) \to \Lambda^{n-p}(M)$, where the orientation of $M$ is fixed by the choice of a volume form $*1= e^0 \wedge e^1 \wedge ... \wedge e^{n-1}$. Finally, a linear connection $\nabla$ on $M$ can be given by a set of connection 1-forms $\{\Lambda^a_{\ b}\}$ so that $\nabla_{X_a} X_b = \Lambda^c_{\ b} (X_a) X_c$. A linear connection $\nabla$ is uniquely fixed by the non-metricity, torsion, and curvature forms defined via the Cartan's structure equations below:
\begin{align}
\overset{(\Lambda)}{D}\eta_{ab}&= -(\Lambda_{ab} + \Lambda_{ba}) = -2Q_{ab},\\
\overset{(\Lambda)}{D}e^a &= de^a + \Lambda^a_{\ b} \wedge e^b = T^a, \\
\overset{(\Lambda)}{D}\Lambda^a_{\ b} &=d\Lambda^a_{\ b} +\Lambda^a_{\ c} \wedge\Lambda^c_{\ b} = \overset{(\Lambda)}{R^a_{\ b}}.
\end{align}
$d$, $\overset{(\Lambda)}{D}$ and $\overset{(\Lambda)}{R^a_{\ b}}$ denote the exterior derivative, exterior covariant derivative and curvature of the above connection, respectively. Bianchi identities are obtained as the integrability conditions of the Cartan's structure equations:
\begin{align}
\overset{(\Lambda)}{D}Q_{ab}&= \frac{1}{2} (\overset{(\Lambda)}{R_{ab}} + \overset{(\Lambda)}{R_{ba}}), \label{bi1}\\
\overset{(\Lambda)}{D}T^a &= \overset{(\Lambda)}{R^a_{\ b}} \wedge e^b, \label{bi2}\\
\overset{(\Lambda)}{D}\overset{(\Lambda)}{R^a_{\ b}} &=0. \label{bi3}
\end{align}

\noindent To see that a generic linear connection is fixed uniquely by the metric tensor field $g$, the torsion tensor field $T$ and a non-metricity tensor field $S=\overset{(\Lambda)}{D} g$, we separate the anti-symmetric and symmetric parts of the connection 1-forms as follows: 
\begin{equation}
\Lambda^a_{\ b} = \Omega^a_{\ b} + Q^a_{\ b},
\end{equation}
where the anti-symmetric part further decomposes in a unique way according to
\begin{equation}
\Omega^a_{\ b} = \omega^a_{\ b} + K^a_{\ b} + q^a_{\ b}.
\end{equation}
Here, the Levi-Civita connection 1-forms $\{\omega^a_{\ b}\}$ are determined completely by the co-frames from the Cartan structure equations
\begin{equation}
de^a+  \omega^a_{\ b} \wedge e^b = 0.
\end{equation}
The contortion 1-forms $\{K^a_{\ b}\}$ are fixed by the torsion 2-forms
\begin{equation}
K^a_{\ b}\wedge e^b = T^a .
\end{equation}
The anti-symmetric 1-forms $\{q^a_{\ b}\}$ are completely determined in terms of the symmetric non-metricity 1-forms $\{Q^a_{\ b}\}$ by the equations
\begin{equation}
q^a_{\ b}=-(\iota^a Q_{bc}) e^c + (\iota_b Q^a_{\ c}) e^c .
\end{equation}

\noindent In a Weyl covariant theory, field elements are allowed to carry some representation of the scale group. The transformation properties of these fields are intimately connected to the dimensions that they are carrying. Under a local scale transformation, a field $\Phi$ transforms as  
\begin{equation}
\Phi \mapsto \exp(-q\sigma) \Phi,
\end{equation}
where $\sigma$ is a dimensionless real scalar field on space-time and the dimensionless parameter $q$ is called the Weyl charge. Conventionally, the metric tensor is assigned a Weyl charge of $-2$ because it has dimension length squared. After that, Weyl charge assignments of other fields are done accordingly. For the linear connection, we adopt the Weyl transformation rule under Weyl group action:
\begin{equation}
\nabla \mapsto \nabla.
\end{equation}
This is a consistent choice because the connection is not a tensorial quantity and therefore is not assigned any dimensions, and therefore stays inert under local scale transformations. Also in a RCW space-time, when at least one of the torsion or non-metricity tensors, or both are present, there need not be any correlations between metric scaling and transformation of the linear connection. \\

\noindent Under a local change of scale, space-time exterior covariant derivative does not transform covariantly. Thus, we introduce a Weyl connection 1-form $Q$ as a compensating potential. $Q$ is a dimensionless 1-form that transforms as 
\begin{equation}
Q \mapsto Q + d\sigma 
\end{equation}
under a local scale transformation. With the help of $Q$, the exterior Weyl covariant derivative of a p-form $\Phi^p_q$ with Weyl charge $q$ is defined as:
\begin{equation}
\mathcal{D} \Phi^p_q = \overset{(\Lambda)}{D} \Phi^p_q + qQ \wedge \Phi^p_q    \label{wcd}
\end{equation}
so that under a local scale transformation $\mathcal{D} \Phi^p_q$ transforms covariantly. Under the action of interior product and Hodge duality operator, Weyl charge of the fields change as:
\begin{align}
\iota_a \Phi^p_q &= \Phi^{p-1}_{q+1}, \\
* \Phi^p_q &= \Phi^{n-p}_{q - (n-2p)}.
\end{align}
To discuss locally Weyl covariant theories in RCW space-times, we take the following relation between the Weyl connection 1-form $Q$ and the the non-metricity tensor $S=\overset{(\Lambda)}{D} g$:
\begin{equation}
\mathcal{D}g=S-2Q\wedge g =0.
\end{equation}
Therefore, the non-metricity 1-forms $\{Q_{ab}\}$ and the Weyl connection 1-form $Q$ are related to each other by:
\begin{equation}
Q_{ab} = -Q \eta_{ab} \label{nonmet}.
\end{equation}
This identification gives a geometrical origin to the Weyl connection and assignment of units to dimensioned quantities.  \\

\noindent The Ricci 1-forms are obtained by contracting the curvature 2-forms:
\begin{equation}
\overset{(\Lambda)}{Ric_a} = \iota_b \overset{(\Lambda)}{R^b_{\ a}}. 
\end{equation}
The curvature scalar needs one more contraction with the metric itself: 
\begin{equation}
 \overset{(\Lambda)}{R} = \iota^a \overset{(\Lambda)}{Ric_a} = \iota^{ab} \overset{(\Lambda)}{R_{ba}}.
\end{equation}
Moreover, the Einstein $(n-1)$-forms of our non-Riemannian connection are defined through the variation of Einstein-Hilbert term as:
\begin{equation}
\overset{(\Lambda)}{G_a}= \overset{(\Lambda)}{G_{ab}} *e^b = -\frac{1}{2} \overset{(\Lambda)}{R_{bc}} \wedge * e^{abc} . \label{efor}
\end{equation}
We note that, although the curvature 2-forms may depend both on the anti-symmetric and symmetric parts, similar to the Riemannian case only the anti-symmetric part of the connection contributes to the Einstein tensor. \\

\noindent We will discuss our gravitational models using an action principle. The field equations are going to be derived using a first order variational formalism. For a non-scale covariant gravitational model, the action functional depends on the co-frame 1-forms $\{e^a\}$, anti-symmetric part of connection 1-forms $\{ \Omega^a_{\ b} = \omega^a_{\ b} + K^a_{\ b} \}$ and possibly on some Lagrange multiplier valued forms $\{\lambda_a\}$. To obtain a scale covariant generalisation, we introduce two more independent variables to original theories: the dilaton 0-form $\alpha$ with the dimension of inverse length and the Weyl connection 1-form $Q$. Then, for a locally scale covariant model, the action functional is given by
\begin{equation}
I[e^a, \Omega^a_{\ b}, Q^a_{\ b}, \alpha, \lambda_a ] = \int_M \mathcal{L}_W \nonumber
\end{equation}
where $\mathcal{L}_W$ is a scale invariant Lagrangian density and $M$ is a compact region in a RCW manifold without boundary. \\

\noindent To check the consistency of the scale covariant generalisation, we use the following diagram:
\begin{center}
\begin{tikzcd}[column sep=1in, row sep=1in]
\mathcal{L} \arrow[r, "\text{introduce }\alpha\ \& \ Q"] \arrow[d, "\text{variation}"'] & \mathcal{L}_W \arrow[d, "\text{variation}"] \\
\dot{\mathcal{L}}  \arrow[r, leftarrow, "\text{set }\alpha=1 \ \&\ Q=0"] &  \dot{\mathcal{L}}_W 
\end{tikzcd}
\end{center}
We introduce scale invariant terms to the Lagrangian $\mathcal{L}$ of the original theory using the dilaton field $\alpha$ and Weyl connection 1-form $Q$. Then vary the scale invariant Lagrangian $\mathcal{L}_W$ and obtain the scale covariant variational field equations $\dot{\mathcal{L}}_W$. If these field equations agree with the field equations of original theory $\dot{\mathcal{L}}$ for a fixed scale $\alpha=1$, and vanishing non-metricity $Q=0$, the above diagram commutes and we say that the generalisation is consistent. \\

\noindent A consistent generalisation means that the scale covariant theory contains the original theory in its vacuum configuration for the Weyl sector. The vacuum configuration means the Weyl connection 1-form has a vanishing field strength, i.e. it is flat. In this case, any solution of the original theory defines an equivalence class of solutions for the scale covariant theory. In this class, two solutions are related to each other by a pure gauge transformation. \\

\section{Quadratic Curvature Gravity in Three Dimensions}

\noindent We start with the formulation of quadratic curvature gravity in a 3D pseudo-Riemannian setting in the language of differential forms and a first order variational formalism. The independent variables are the co-frame 1-forms $\{e^a\}$ and connection 1-forms $\{ \Omega^a_{\ b} \}$. Lagrange multiplier valued 1-forms $\{\lambda_a\}$ are introduced to constraint the space-time torsion to zero. We consider the Lagrangian density 3-form:
\begin{equation}
\mathcal{L}= \frac{1}{K} \overset{(\Omega)}{R^a_{\ b}} \wedge *e_a^{\ b} + \Lambda *1 + \lambda_a \wedge T^a + \kappa_1 \overset{(\Omega)}{R^a_{\ b}} \wedge * \overset{(\Omega)}{R_a^{\ b}} + \kappa_2 \overset{(\Omega)}{Ric^a} \wedge * \overset{(\Omega)}{Ric_a} +\kappa_3 \overset{(\Omega)}{R^2}*1. \label{qcg1}
\end{equation}
where $K$ denotes the three dimensional gravitational constant, $\Lambda$ the cosmological constant and $\kappa_1$, $\kappa_2$ and $\kappa_3$ are coupling constants with dimensions of inverse length. This family of Lagrangian densities also cover the Lagrangian density of the NMG theory for the particular values of $\kappa_1=0$, $\kappa_2=1$ and $\kappa_3=-3/8$. \\

\noindent In our formulation, we will vary the metric compatible totally anti-symmetric connection 1-forms:
\begin{equation}
\Omega^a_{\ b}= \omega^a_{\ b}+K^a_{\ b}. \label{con}
\end{equation}

\noindent The total variational derivative of $\mathcal{L}$ with respect to three independent variables is found to be:
\begin{align}
\dot{\mathcal{L}} &= {\dot{e}}^a \wedge \bigg\{ \frac{1}{K} \overset{(\Omega)}{R^b_{\ c}} \epsilon_{ab}^{\ \ c} +\Lambda *e_a + \overset{(\Omega)}{D}\lambda_a -\kappa_1 \hat{\tau_a}[\overset{(\Omega)}{R^b_{\ c}}]\nonumber\\
&+\kappa_2[\iota_a(\overset{(\Omega)}{Ric^b} \wedge * \overset{(\Omega)}{Ric_b})+2\iota_a \overset{(\Omega)}{R_{bc}}\wedge\iota^b*\overset{(\Omega)}{Ric^c}]+\kappa_3(2 \overset{(\Omega)}{R}\overset{(\Omega)}{R^b_{\ c}} \epsilon_{ab}^{\ \ c}-\overset{(\Omega)}{R^2}*e_a)\bigg\} \nonumber\\
&+\dot{\Omega^a_{\ b}} \wedge \bigg\{\overset{(\Omega)}{D}*e_a^{\ b}+e^b \wedge \lambda_a +2 \overset{(\Omega)}{D}[\kappa_1*\overset{(\Omega)}{R_a^{\ b}}-\kappa_2 \iota_a *\overset{(\Omega)}{Ric^b}+\kappa_3R*e_a^{\ b}] \bigg\}\nonumber\\
& + \dot{\lambda_a}\wedge T^a.
\end{align}
Above, a dot over a field variable denotes the variation of that variable. First, due to Lagrange constraint equation, torsion vanishes and we will be working with the unique Levi-Civita connection 1-forms $\{\omega^a_{\ b}\}$. Then, we solve the Lagrange multiplier 1-forms from the connection variation equation. For this, we write the connection equation as:
\begin{equation}
\frac{1}{2}(e_a \wedge \lambda_b - e_b \wedge \lambda_a) =\Sigma_{ab},
\end{equation}
where
\begin{equation}
\Sigma_{ab}= 2 \overset{(\omega)}{D} (\kappa_1* \overset{(\omega)}{R_{ab}}-\kappa_2 \iota_{[a}*\overset{(\omega)}{Ric_{b]}}+\kappa_3 \overset{(\omega)}{R}*e_{ab}) \nonumber
\end{equation}
is anti-symmetric due to anti-symmetry of the Levi-Civita connection 1-forms. Indices between square brackets means total anti-symmetrization of those indices. Then, the unique solution for the Lagrange multiplier 1-forms reads:
\begin{equation}
\lambda_a=2\iota^b\Sigma_{ba}-\frac{1}{2}(\iota^{bc}\Sigma_{cb})e_a.
\end{equation}
Finally, Einstein field equations are determined to be:
\begin{align}
-\frac{2}{K}& \overset{(\omega)}{G_a}+\Lambda*e_a+\overset{(\omega)}{D}\lambda_a-\kappa_1\hat{\tau}[\overset{(\omega)}{R^b_{\ c}}]+\kappa_3(2 \overset{(\omega)}{R}\overset{(\omega)}{R^b_{\ c}} \epsilon_{ab}^{\ \ c}-\overset{(\omega)}{R^2}*e_a) \nonumber\\
&+\kappa_2[\iota_a(\overset{(\omega)}{Ric^b} \wedge * \overset{(\omega)}{Ric_b})+2\iota_a \overset{(\omega)}{R_{bc}}\wedge\iota^b*\overset{(\omega)}{Ric^c}]=0. \label{qcgefe}
\end{align}

\noindent In order to promote above model into a locally scale covariant one, we introduce two new independent variables: dilaton scalar $\alpha$ and the Weyl 1-form $Q$, and consider the most general connection 1-forms that have their symmetric part identified with the Weyl connection 1-form
\begin{equation}
\Lambda^a_{\ b}=\omega^a_{\ b}+K^a_{\ b}+q^a_{\ b}-Q\eta^a_{\ b}.
\end{equation}
Thus, we consider the following Weyl invariant Lagrangian density 3-form\footnote{ In our earlier papers \cite{dereli-yetismisoglu1, dereli-yetismisoglu2}, using an identy we replaced one of the three quadratic curvature invariants in terms of the other two. Here, we do not do this and keep all three quadratic curvature invariants. We explain this choice and the derivation of the identity in appendix A.}: 
\begin{align}
\mathcal{L}_W&=\alpha \overset{(\Lambda)}{R^a_{\ b}} \wedge *e_a^{\ b} + \alpha^3\Lambda*1 +\alpha\lambda_a\wedge T^a-\frac{\gamma}{2\alpha} \mathcal{D} \alpha \wedge * \mathcal{D} \alpha -\frac{\gamma^{\prime}}{2\alpha}dQ \wedge *dQ   \nonumber\\
& + \frac{1}{\alpha}\bigg[ \kappa_1 \overset{(\Lambda)}{R^a_{\ b}} \wedge * \overset{(\Lambda)}{R_a^{\ b}} + \kappa_2 \overset{(\Lambda)}{Ric^a} \wedge * \overset{(\Lambda)}{Ric_a} + \kappa_3 \overset{(\Lambda)}{R^2}*1 \bigg]. \label{lw}
\end{align}
Above, to promote $\alpha$ and $Q$ to dynamical fields, we added their kinetic terms where $\gamma$ and $\gamma'$ are new dimensionless coupling constants. While finding variational field equtions, we vary the Lagrangian density with respect to the total connection 1-forms $\{\Lambda^a_{\ b}\}$. Then, we will separate the connection variation equations acccording to 
\begin{equation}
\dot{\Lambda^a_{\ b}}=\dot{\Omega^a_{\ b}}-\eta^a_{\ b} \dot{Q}. \label{sep}
\end{equation}

\noindent Therefore the variation of the Lagrangian denstiy (\ref{lw}) is found to be:
\begin{align}
\dot{\mathcal{L}}_W &= {\dot{e}}^a \wedge \bigg\{ \alpha \overset{(\Lambda)}{R^b_{\ c}} \epsilon_{ab}^{\ \ c} +\alpha^3\Lambda *e_a + \overset{(\Lambda)}{D}(\alpha\lambda_a)+\frac{\gamma}{2\alpha}\tau_a[\mathcal{D}\alpha]+\frac{\gamma}{2\alpha}\hat{\tau}_a[dQ] \nonumber\\
&-\frac{\kappa_1}{\alpha}\hat{\tau}_a[\overset{(\Lambda)}{R^b_{\ c}}] + \frac{\kappa_2}{\alpha} \bigg[\iota_a\big( \overset{(\Lambda)}{Ric_b} \wedge * \overset{(\Lambda)}{Ric^b} \big) +2\iota_a \overset{(\Lambda)}{R_{bc}} \wedge \iota^b * \overset{(\Lambda)}{Ric^c} \bigg]  \nonumber\\
&+\frac{\kappa_3}{\alpha} \bigg[ 2\overset{(\Lambda)}{R} \overset{(\Lambda)}{R^b_{\ c}}\epsilon_{ab}^{\ \ c}-\overset{(\Lambda)}{R^2} *e_a \bigg]\bigg\}+ \dot{\lambda_a} \wedge \big(\alpha T^a\big)\nonumber\\
&+ {\dot{\Lambda}}^a_{\ b} \wedge \bigg\{\overset{(\Lambda)}{D} \bigg[\alpha *e_a^{\ b} + \frac{2\kappa_1}{\alpha}*\overset{(\Lambda)}{R_a^{\ b}}-\frac{2\kappa_2}{\alpha}\iota_a*\overset{(\Lambda)}{Ric^b}+\frac{2\kappa_3}{\alpha} \overset{(\Lambda)}{R}*e_a^{\ b} \bigg] \nonumber\\
&+\alpha e^b \wedge \lambda_a + \frac{1}{3} \eta^b_{\ a}\gamma*\mathcal{D}\alpha+ \frac{1}{3} \eta^b_{\ a}  \gamma'd\bigg(\frac{1}{\alpha}*dQ \bigg) \bigg\} \nonumber\\
&+\dot{\alpha}\bigg\{\overset{(\Lambda)}{R^a_{\ b}} \wedge *e_a^{\ b} +3\alpha^2 \Lambda*1 +\frac{\gamma}{2\alpha^2}\mathcal{D}\alpha \wedge*\mathcal{D}\alpha+\gamma\mathcal{D}\bigg(\frac{1}{\alpha}*\mathcal{D}\alpha\bigg) +\lambda_a \wedge T^a \nonumber\\
&+\frac{\gamma'}{2\alpha^2}dQ \wedge *dQ-\frac{1}{\alpha^2}\bigg[\kappa_1 \overset{(\Lambda)}{R^a_{\ b}} \wedge * \overset{(\Lambda)}{R_a^{\ b}} + \kappa_2 \overset{(\Lambda)}{Ric^a} \wedge * \overset{(\Lambda)}{Ric_a} + \kappa_3 \overset{(\Lambda)}{R^2}*1 \bigg]\bigg\}.
\end{align}
We start simplifying by first noting that the torsion 2-forms vanish. Then, we first go to dilaton field equation. To do this, we compare the trace of the co-frame equations
\begin{align}
e^a \wedge \frac{\delta \mathcal{L}_W}{\delta e^a} &= \alpha\overset{(\Lambda)}{R^a_{\ b}} \wedge *e_a^{\ b} +3\alpha^3 \Lambda*1 -\frac{\gamma}{2\alpha}\mathcal{D}\alpha \wedge*\mathcal{D}\alpha+\frac{\gamma'}{2\alpha}dQ \wedge*dQ \nonumber\\
&+\alpha\lambda_a \wedge T^a-d(\alpha e^a \wedge \lambda_a)- \frac{\kappa_1}{\alpha}\overset{(\Lambda)}{R^a_{\ b}} \wedge * \overset{(\Lambda)}{R_a^{\ b}} +\frac{3\kappa_2}{\alpha}\overset{(\Lambda)}{Ric^a} \wedge * \overset{(\Lambda)}{Ric_a}\nonumber\\
& +\frac{4\kappa_2}{\alpha}\overset{(\Lambda)}{R^a_{\ b}} \wedge \iota^a * \overset{(\Lambda)}{Ric^b} - \frac{\kappa_3}{\alpha}\overset{(\Lambda)}{R^2}*1=0,
\end{align}
with the dilaton field equation above and obtain: 
\begin{equation}
d(\alpha e_a \wedge \lambda^a+\gamma*\mathcal{D}\alpha)=0. \label{lw1}
\end{equation}
Next, we separate the symmetric and anti-symmetric parts of the connection variaton equations by lowering an index. In order to lower an index inside a covariant derivative, we make use of the following identities: 
\begin{equation}
\overset{(\Lambda)}{D}(\alpha*e_a^{\ b})= \mathcal{D}\alpha \wedge *e_a^{\ b}, \label{ide1}
\end{equation}
\begin{equation}
\overset{(\Lambda)}{D}\bigg(\frac{1}{\alpha}*\overset{(\Lambda)}{R_a^{\ b}}\bigg)=-\frac{2Q}{\alpha} \wedge  *\overset{(\Lambda)}{R_a^{\ b}} -\frac{d\alpha}{\alpha^2} \wedge  *\overset{(\Lambda)}{R_a^{\ b}} + \frac{1}{\alpha}\eta^{cb}\overset{(\Lambda)}{D}*\overset{(\Lambda)}{R_{ac}}, \label{ide2}
\end{equation}
\begin{equation}
\overset{(\Lambda)}{D}\bigg(\frac{1}{\alpha}\iota_a*\overset{(\Lambda)}{Ric^b}\bigg)= -\frac{2Q}{\alpha} \wedge\iota_a*\overset{(\Lambda)}{Ric^b} -\frac{d\alpha}{\alpha^2} \wedge \iota_a*\overset{(\Lambda)}{Ric^b}+ \frac{1}{\alpha}\eta^{cb}\overset{(\Lambda)}{D}(\iota_a*\overset{(\Lambda)}{Ric_c}), \label{ide3}
\end{equation}
\begin{equation}
\overset{(\Lambda)}{D}\bigg(\frac{1}{\alpha}\overset{(\Lambda)}{R}*e_a^{\ b}\bigg)=\frac{Q}{\alpha} \wedge \overset{(\Lambda)}{R}*e_a^{\ b}+\frac{1}{\alpha}d\overset{(\Lambda)}{R}\wedge*e_a^{\ b}-   \frac{d\alpha}{\alpha^2} \wedge \overset{(\Lambda)}{R}*e_a^{\ b}. \label{ide4}
\end{equation}
On the right hand side of identities (\ref{ide1}) - (\ref{ide4}) there are terms proportional to the torsion 2-forms in general, however that should be omitted here as they identically vanish. \\

\noindent After lowering an index and using (\ref{sep}), the symmetric and anti-symmetric parts of the connection variation equations read:
\begin{equation}
\alpha e^a \wedge \lambda_a +\gamma*\mathcal{D}\alpha= (6\kappa_1+6\kappa_2-\gamma')d\bigg(\frac{1}{\alpha}*dQ\bigg), \label{lw2}
\end{equation}
and 
\begin{align}
\frac{\alpha}{2}(&e_a\wedge\lambda_b - e_b\wedge\lambda_a)=\Sigma_{ab},  \label{antisymw}
\end{align}
respectively, where
\begin{align}
\Sigma_{ab}&=\bigg[\mathcal{D}\alpha+\frac{2\kappa_3}{\alpha}\bigg(d\overset{(\Omega)}{R}+Q\overset{(\Omega)}{R}-\frac{d\alpha}{\alpha}\overset{(\Omega)}{R}\bigg)\bigg]\wedge*e_{ab}+2\kappa_1\overset{(\Omega)}{D}\bigg(\frac{1}{\alpha}*\overset{(\Omega)}{R_{ab}}\bigg)\nonumber\\
&+2\kappa_2\overset{(\Omega)}{D}\bigg(\frac{1}{\alpha}\iota_{[a}*\iota_{b]}dQ-\frac{1}{\alpha}\iota_{[a}*\overset{(\Omega)}{Ric_{b]}}\bigg). \nonumber
\end{align} 
 When writing the anti-symmetric part of the equations, we used the fact that index raising and lowering operations commute with the covariant derivative operation with respect to the anti-symmetric connection 1-forms $\{\Omega^a_{\ b}\}$. We algebraically solve (\ref{antisymw}) for the Lagrange multiplier 1-forms as:
\begin{equation}
\lambda_a=\frac{2}{\alpha}\iota^b\Sigma_{ba}-\frac{1}{2\alpha}(\iota^{bc}\Sigma_{cb})e_a. \label{lw3}
\end{equation}
The substitution of above in Einstein field equations of the Weyl covariant theory gives:
\begin{align}
&-2\alpha\overset{(\Omega)}{G_a}+\alpha^3\Lambda*e_a+\overset{(\Omega)}{D}(\alpha\lambda_a)+\alpha Q\wedge \lambda_a+\frac{\gamma}{2\alpha}\tau_a[\mathcal{D}\alpha]+\bigg(\frac{\gamma'}{2\alpha}-\frac{3\kappa_1}{\alpha}\bigg)\hat{\tau}_a[dQ] \nonumber\\
& - \frac{\kappa_1}{\alpha}\hat{\tau}_a[\overset{(\Omega)}{R^b_{\ c}}]+ \frac{\kappa_2}{\alpha} \bigg[\iota_a\big( \overset{(\Omega)}{Ric_b} \wedge * \overset{(\Omega)}{Ric^b} \big) +2\iota_a \overset{(\Omega)}{R_{bc}} \wedge \iota^b * \overset{(\Omega)}{Ric^c} +4\iota_a(dQ\wedge*dQ) \nonumber\\
&-2\iota_a \overset{(\Omega)}{R_{bc}} \wedge\iota^b*\iota^c dQ -6\iota_a dQ \wedge *dQ \bigg]+\frac{\kappa_3}{\alpha} \bigg[ 2\overset{(\Omega)}{R} \overset{(\Omega)}{R^b_{\ c}}\epsilon_{ab}^{\ \ c}-\overset{(\Omega)}{R^2} *e_a \bigg]=0. \label{lw4}
\end{align}
Therefore the Weyl covariant quadratic curvature theory defined via the action (\ref{lw}) yields three sets of field equations: (\ref{lw1}), (\ref{lw2}), and (\ref{lw4}). In order to show the consistency of this generalisation, we are going to restrict the Weyl sector of the theory to its vacuum sector and show that the originial quadratic curvature theory field equations are contained in this configuration. To this end, we make the following choices: 
\begin{equation}
\mathcal{D}\alpha=0, \quad \alpha=1 \ \ \implies \ \ Q=0. \label{elim}
\end{equation} 
The first choice sets the Weyl connection to be a flat connection, then the second choice fixes a global units scale. Consequently, the Weyl connection 1-form gets cancelled out and we are left with a pseudo-Riemannian geometry. Then the field equations reduce to:
\begin{align}
-2& \overset{(\omega)}{G_a}+\Lambda*e_a+\overset{(\omega)}{D}\lambda_a-\kappa_1\hat{\tau}[\overset{(\omega)}{R^b_{\ c}}]+\kappa_3(2 \overset{(\omega)}{R}\overset{(\omega)}{R^b_{\ c}} \epsilon_{ab}^{\ \ c}-\overset{(\omega)}{R^2}*e_a) \nonumber\\
&+\kappa_2[\iota_a(\overset{(\omega)}{Ric^b} \wedge * \overset{(\omega)}{Ric_b})+2\iota_a \overset{(\omega)}{R_{bc}}\wedge\iota^b*\overset{(\omega)}{Ric^c}]=0.  \label{red4}
\end{align}
and 
\begin{equation}
e^a \wedge \lambda_a=0, \qquad d(e^a \wedge \lambda_a)=0  \label{red1}
\end{equation}
where
\begin{align}
\lambda_a&=2\iota^b\Sigma_{ba}-\frac{1}{2}(\iota^{bc}\Sigma_{cb})e_a, \quad \text{and}\label{red3} \nonumber\\
\Sigma_{ab}&=2 \overset{(\omega)}{D} (\kappa_1* \overset{(\omega)}{R_{ab}}-\kappa_2 \iota_{[a}*\overset{(\omega)}{Ric_{b]}}+\kappa_3 \overset{(\omega)}{R}*e_{ab}).   \nonumber
\end{align}
Although the equations (\ref{red4}) agree with the field equations (\ref{qcgefe}) of quadratic curvature gravity, the equations (\ref{red1}) are extra. One must make sure that the equation (\ref{red1}) vanishes\footnote{This guarantees the equation $d(e^a\wedge \lambda_a)=0$ is also satisfied, and we are only left with quadratic curvature gravity field equations.} identically, so that the Weyl covariant generalisation is consistent. \\

\noindent The explicit evaluation of equation (\ref{red1}) is technical and we present it separately in the appendix B. We show that: 
\begin{equation}
e^a \wedge \lambda_a =0 \quad  \Leftrightarrow \quad (2\kappa_1+3\kappa_2+8\kappa_3)\iota_a*d\overset{(\omega)}{R}=0. \label{coneq}
\end{equation}
Then, either the space-time has constant curvature or else in a generic space-time, only certain combinations of quadratic curvature invariants for which 
\begin{equation}
2\kappa_1+3\kappa_2+8\kappa_3=0 \label{curvcons}
\end{equation}
are allowed. It is remarkable that NMG meets this condition: $\kappa_1=0, \kappa_2=1, \kappa_3=-\frac{3}{8}.$

\section{Concluding Remarks}

\noindent We have derived the locally scale covariant extension of quadratic curvature gravity field equations in 3-dimensional Riemann-Cartan-Weyl space-times. Our basic field variables are the space-time metric $g$, the dilaton 0-form $\alpha$ and the Weyl potential 1-form $Q$. Their field equations are obtained by a first order variational principle from a locally scale invariant action in which the space-time torsion is constrained to zero by the method of Lagrange multipliers.  The locally scale covariant variational equations that follow from action density that involves the Einstein-Hilbert term with a cosmological constant plus the most general linear combination of  quadratic curvature invariants in 3-dimensions. We also discussed the consistency of the conformal equivalence class of the vacuum configuration of the theory that is determined by setting the Weyl potential to zero ($Q=0$) and fixing a scale (by the choice $\alpha=1$). We noted that such a consistency imposes a condition on the choice of the quadratic curvature invariants allowed in the action. Rather than a 2-parameter family of an action density 3-form that one would expect, only a 1-parameter family is allowed. It is remarkable that the NMG action belongs to this family.\footnote{A Weyl covariant generalisation of NMG  and quadratic curvature gravity in general has also been studied in \cite{dengiz-tekin}
from a Higgs-like symmetry breaking point of view for NMG. They pointed out in a linearised approximation that  gravitons thus gain mass in ADS and/or Minkowski backgrounds.} \\

\noindent For future direction of research, one may consider to investigate linearised field equations \cite{cebeci-sarioglu-tekin, baykal-dereli} to determine the particle spectrum of the theory and to check the linearisation instabilities \cite{altas-tekin}. Besides, finding out non-trivial solutions that have non-vanishing field strength for the Weyl sector would be interesting. Furthermore, one can look for solutions that are Einstein-Weyl spaces. Due to their specific geometrical properties \cite{pedersen-tod}, 3-dimensional Einstein-Weyl spaces can be formulated in terms of mini-twistor spaces \cite{jones-tod} and can be used to construct four dimensional self-dual geometries \cite{hitchin}. Finally, similar methods can be applied to the NIMG model so that we have a generic Weyl covariant theory that contains important 3D models such as TMG, MMG and NMG altogether.

\newpage

\appendix

\section{Quadratic Curvature Invariants}

\noindent The curvature 2-forms can be uniquely decomposed into their symmetric and anti-symmetric parts according to
\begin{equation}
\overset{(\Lambda)}{R^{a}_{\;\;b}} = \overset{(\Omega)}{R^{a}_{\;\;b}} -\eta^{a}_{\;\;b} dQ.
\end{equation}
Then we also have by contractions
\begin{equation}
\overset{(\Lambda)}{Ric}_{a} = \overset{(\Omega)}{Ric}_{a} - \iota_{a} dQ , \quad 
\overset{(\Lambda)}{R}=\overset{(\Omega)}{R}.
\end{equation}
The co-frame variations of the Einstein-Hilbert term in the action density (3.6) give
\begin{equation}
-\frac{1}{2} \overset{(\Omega)}{R^{bc}} \epsilon_{abc} = *{\overset{(\Omega)}{Ric_a}} -\frac{1}{2}\overset{(\Omega)}{R} *e_a 
\end{equation}
that can be inverted in $n=3$ dimensions as 
\begin{equation}
\overset{(\Omega)}{R^{bc}} = -\epsilon^{abc} \left ( *{\overset{(\Omega)}{Ric_a}} -\frac{1}{2}\overset{(\Omega)}{R} *e_a  \right ).
\end{equation}
Squaring both sides and simplifying we arrive at the following identity satisfied by the quadratic curvature invariants in $n=3$ dimensions: 
\begin{equation}
\overset{(\Omega)}{R^{a}_{\;\;b}} \wedge *\overset{(\Omega)}{R_{a}^{\;\;b}} = 2 \overset{(\Omega)}{Ric_a} \wedge *{\overset{(\Omega)}{Ric^a}} - \frac{1}{2} \overset{(\Omega)}{{R}^2} *1.
\end{equation}
On the other hand from above we also have
\begin{equation}
\overset{(\Lambda)}{R^{a}_{\;\;b}} \wedge *\overset{(\Lambda)}{R_{a}^{\;\;b}} = \overset{(\Omega)}{R^{a}_{\;\;b}} \wedge *\overset{(\Omega)}{R_{a}^{\;\;b}} + 3 dQ \wedge *dQ 
\end{equation}
and
\begin{equation}
\overset{(\Lambda)}{Ric_a} \wedge *{\overset{(\Lambda)}{Ric^a}}=  \overset{(\Omega)}{Ric_a} \wedge *{\overset{(\Omega)}{Ric^a}} + 2 dQ \wedge *dQ -2 \overset{(\Omega)}{Ric^a} \wedge *\iota_{a} dQ.
\end{equation}
In order to simplify the third term on the right hand side, we consider the second Bianchi identity written in the form
\begin{equation}
\overset{(\Omega)}{R^{a}_{\;\;b}} \wedge e^b - dQ \wedge e_a = \overset{(\Lambda)}{D} T^a
\end{equation}
and contract on both sides to get
\begin{equation}
\overset{(\Omega)}{Ric_a} \wedge e^a = dQ + \iota_{a}( \overset{(\Lambda)}{D} T^a ).
\end{equation}
Therefore
\begin{equation}
-2 \overset{(\Omega)}{Ric^a} \wedge *\iota_{a} dQ = 2 \overset{(\Omega)}{Ric_a} \wedge e^{a} \wedge *dQ = 2 dQ \wedge *dQ +2 \iota_{a}( \overset{(\Lambda)}{D} T^a ) \wedge *dQ.
\end{equation}
Putting all these back into our basic quadratic curvature identity above,  we may write it as
\begin{equation}
\overset{(\Lambda)}{R^{a}_{\;\;b}} \wedge *{\overset{(\Lambda)}{R_{a}^{\;\;b}}} - 2\overset{(\Lambda)}{Ric}_{a} \wedge *\overset{(\Lambda)}{Ric^{a}} + \frac{1}{2} {\overset{(\Lambda)}{R^2}} *1= -5 dQ \wedge *dQ -4 \iota_{a}( \overset{(\Lambda)}{D} T^a ) \wedge *dQ. 
\end{equation}

\section{Derivation of Equation (\ref{red1})}
Before starting to calculate (\ref{red1}), we note that
\begin{equation}
e^a \wedge \lambda_a = 2 e^a \wedge \iota^b\Sigma_{ba} = 2 \iota^b(e^a \wedge \Sigma_{ab}).
\end{equation}
Now, we will play with the anti-symmetric 1-form $\Sigma_{ab}$. First using $\iota_a\chi=(-1)^p*(e_a \wedge *\chi)$ for a $p$-form $\chi$, we see    
\begin{equation}
\iota_a*\overset{(\omega)}{Ric_b}= *(\overset{(\omega)}{Ric_b} \wedge e_a) \label{id9}
\end{equation}
where we also made use of $**\chi=-\chi$ for a 2-form $\chi$. Using (\ref{id9}), $\Sigma_{ab}$ can be written as:
\begin{equation}
\Sigma_{ab}=2\overset{(\Omega)}{D}*\bigg[\kappa_1 \overset{(\omega)}{R_{ab}}+\frac{\kappa_2}{2}(e_a \wedge \overset{(\omega)}{Ric_b}-e_b \wedge \overset{(\omega)}{Ric_a})+\kappa_3 \overset{(\omega)}{R} e_{ab}  \bigg]
\end{equation}
Then using the fact that geometry is torsion free, we can write
\begin{equation}
e^a \wedge \Sigma_{ab}=-2\overset{(\Omega)}{D}\bigg[e^a \wedge * \bigg( \kappa_1 \overset{(\omega)}{R_{ab}}+\frac{\kappa_2}{2}(e_a \wedge \overset{(\omega)}{Ric_b}-e_b \wedge \overset{(\omega)}{Ric_a})+\kappa_3 \overset{(\omega)}{R} e_{ab} \bigg)\bigg]
\end{equation}
Using, $\iota_a\chi=(-1)^p*(e_a \wedge *\chi)$ for a $p$-form $\chi$ implies $e^a \wedge *\chi=-*\iota^a \chi$ and above equation reads:
\begin{align}
2\iota^b(e^a \wedge \Sigma_{ab})&=4 \overset{(\Omega)}{D^b}*\iota^a\bigg[\kappa_1 \overset{(\omega)}{R_{ab}}+\frac{\kappa_2}{2}(e_a \wedge \overset{(\omega)}{Ric_b}-e_b \wedge \overset{(\omega)}{Ric_a})+\kappa_3 \overset{(\omega)}{R} e_{ab}\bigg] \nonumber\\
&= 4 \overset{(\Omega)}{D^a}*\bigg[\bigg(\kappa_1 +\frac{\kappa_2}{2}\bigg)\overset{(\omega)}{Ric_a} + \bigg(2\kappa_3+\frac{\kappa_2}{2}\bigg)\overset{(\omega)}{R}e_a\bigg]. \label{id10}
\end{align}
Finally equating (\ref{id10}) to zero yields the equation (\ref{coneq}).

\end{document}